\documentstyle[version2,preprint,aps]{revtex}
\begin{document}
\draft
\preprint{}
\begin{title}
Nucleation of quark matter bubbles in neutron stars
\end{title}
\author{Michael L.\ Olesen and Jes Madsen}
\begin{instit}
Institute of Physics and Astronomy, Aarhus University,
DK-8000 \AA rhus C, Denmark
\end{instit}
\centerline{To be published in Phys.~Rev.~D}
\receipt{1 July 1993}
\begin{abstract}
The thermal nucleation of quark matter bubbles inside neutron stars is
examined for various temperatures which the star may realistically
encounter during its lifetime. It is found that for a bag constant less
than a critical value, a very large part of the star will be converted into
the quark phase within a fraction of a second. Depending on the equation of
state for neutron star matter and strange quark matter, all or some
of the outer parts of the star may subsequently
be converted by a slower burning or a detonation.
\end{abstract}

\narrowtext
\section{INTRODUCTION}
\label{secintro}

If pulsars or the central parts of these
can be made of quark matter rather than neutrons
\cite{witten84}, does this then apply to all or just some
of them, and when and how does the phase transformation take place?

According to some investigations \cite{horvath}, the transformation
occurs during the supernova explosion. In this scenario, the released
binding
energy is what makes the supernova succeed in the first place, supplying the
final ``push'' which seems to lack in most of the computer simulations of
the events.

Another model for strange star formation (in the context of absolutely stable
strange quark matter) was introduced by Baym {\it et al.}
\cite{olinto}, describing the transformation as a slow burning
(combustion) rather than a violent event connected with a supernova
detonation.

Regardless of the way in which the transformation occurs, an
initial seed of quark matter is needed to start it. Alcock {\it et al.}
\cite{alcock} suggested a variety of possibilities ranging from
pressure induced conversion via two flavor quark matter to collision with
either highly energetic neutrinos or smaller lumps of strange quark matter.
However, they
did not provide a rate for conversion of neutron stars and thus left it as
an open question, whether every compact object is a strange star, or whether
they are rare objects, even if quark matter formation is energetically
favorable.
It has also been suggested \cite{madsenfriedman} that strange matter seeds
(in the case of quark matter stability) from strange star collisions or of
cosmological origin would trigger the transformation of all neutron stars,
in which case the thermal nucleation would be of relevance to the case of
unstable quark matter only \cite{hoenenogaegget}.

Other possibilities are that shock waves in the supernova trigger the
conversion; a seed could be produced by non-thermal quantum
fluctuations, or a phase transition could be started around impurities. We
are not able to estimate the probability of either method,
but would expect at least quantum fluctuations to be less likely than the
thermal nucleation process discussed below.

An estimate for quark matter formed via thermally induced fluctuations
was given by Horvath {\it et al.} \cite{deltamu},
using typical numbers for various physical quantities.
It was found that
all neutron stars are converted into strange stars (assuming stable strange
quark matter) if the temperature at some time during
the stars lifetime has exceeded 2-3 MeV.

In the following, we choose an approach similar to the one in Ref.\
\cite{deltamu}, but with an extra term in the expression for the surface
energy of quark matter, and with the two phases treated in a more
self-consistent way. Unlike most of the approaches mentioned above
\cite{witten84,horvath,olinto,alcock,madsenfriedman,deltamu}, we will be
considering the
formation of both strange stars (for absolutely stable strange matter) and
hybrid stars, where strange matter is formed only in the central regions
due to the high pressure.
First, we will deal with some
general aspects of nucleation (Sec.\ \ref{secboil}). For pedagogical
purposes, Sec.\ \ref{pure}
treats the problem using a simplified model with the hadron phase being a
free degenerate neutron gas and the quark phase a bag model with
only $u$ and $d$ quarks. A more detailed model for the neutron star is
presented
in Sec.\
\ref{seccompli}, followed by some concluding remarks (Sec.\ \ref{secclu}).

\section{BUBBLE FORMATION}
\label{secboil}

The free energy involved in formation of a spherical quark bubble of radius
$R$ is given by
\begin{equation}
F = - {{4 \pi} \over 3} R^3 \Delta P + 4 \pi \sigma R^2  + 8 \pi \gamma R
+ N_q \Delta\mu ,
\label{work}
\end{equation}
where $\Delta P = P_q - P_h$ is the pressure difference,
$\sigma = \sigma_q + \sigma_h$ the surface tension,
$\gamma = \gamma_q - \gamma_h$ the curvature energy density, and $\Delta \mu
= \mu_q - \mu_h$ the difference in chemical potential. $N_q$ is the total
baryon
number in the quark bubble. The indices $h$ and $q$ denote the hadron and
quark phase respectively. (We set $T$ = 0 in the thermodynamical
expressions; since $T \ll \mu$ throughout, this only leads to minor
errors).  Defining $C = C(\mu_q) \equiv \Delta P - n_q
\Delta \mu$ and $b \equiv 2\gamma C/ \sigma^2$ the free energy has a maximum
at the critical radius
\begin{equation}
r_c = {\sigma\over{C}}\left( 1 + \sqrt{1 + b}\right) ,
\label{rcrit}
\end{equation}
and the corresponding free energy
\begin{equation}
W_c \equiv F(r_c) = {{4 \pi} \over 3} {\sigma^3 \over {C^2}} \left[
2 + 2(1+b)^{3/2} + 3b\right]
\label{wcrit}
\end{equation}
is the work required to form a bubble of this radius which is the smallest
bubble capable of growing.

It is a standard assumption in the theory of
bubble nucleation in first order phase transitions that bubbles form at
this particular radius at a rate given by \cite{prefactor}

\begin{equation}
{\cal R} \approx T^4 \exp (-W_c/T) .
\label{prob}
\end{equation}

It is possible to show that $W_c$ has a minimum as a function of $\mu_q$ at
$\mu_q = \mu_h$. This gives a maximum in the rate for bubble formation, and
because of the exponential in Eq.\ (\ref{prob}) one may safely assume that
nucleation happens in chemical equilibrium. Thus, $C$ reduces to $C = \Delta
P$.

Throughout this paper, we will consider the strange quark to be massless,
in which case chemical equilibrium would give
$\mu_u = \mu_d = \mu_s = {1\over3} \mu_h$.
However, equilibrium is established
only on a weak interaction time scale, whereas the formation of bubbles is
governed by the strong interaction and is many orders of magnitude faster.
So, although we have chemical equilibrium between the two phases this is
not so between the three quark flavors. Instead, flavor must be
conserved during the phase transition.

A consequence of having massless quarks is that $\sigma_q \equiv 0$
\cite{bergerjaffe}, and
since $\sigma_h$ is negligible compared to $\gamma$, Eqs.\ (\ref{rcrit}) and
(\ref{wcrit}) reduce to

\begin{equation}
r_c = \sqrt{{{2\gamma}\over{\Delta P}}}
\label{rcrit2}
\end{equation}
and

\begin{equation}
W_c = {{16\pi}\over 3} \sqrt{{{2\gamma^3}\over{\Delta P}}} .
\label{wcrit2}
\end{equation}

\section{PURE NEUTRON GAS}
\label{pure}

Before considering a more realistic equation of state it is instructive
to study the nucleation of a pure neutron gas into quarks. The
quark bubbles formed consist of $u$ and $d$ quarks in the ratio 1:2;
only later weak interactions may change the composition to an
energetically more favorable state. Thus quark chemical potentials are
related by $\mu_d=2^{1/3}\mu_u$, and
$\mu_n=\mu_u+2\mu_d=(1+2^{4/3})\mu_u$, assuming chemical equilibrium
across the phase boundary.

The pressure difference is given by

\begin{equation}
\Delta P=P_{ud}-P_n= {{\mu_u^4+\mu_d^4}\over{4\pi^2}}-B-P_n
\label{deltaP}
\end{equation}
and the curvature energy coefficient \cite{madsen}

\begin{equation}
\gamma={{\mu_u^2+\mu_d^2}\over{8\pi^2}} .
\label{gamma}
\end{equation}

For the question in hand we choose the simplest possible equation of
state for the neutron gas, namely that of a zero temperature,
nonrelativistic degenerate Fermi-gas, where

\begin{equation}
P_n={{(\mu_n^2-m_n^2)^{5/2}}\over{15\pi^2m_n}}
\end{equation}
and the baryon density

\begin{equation}
n_B={{(\mu_n^2-m_n^2)^{3/2}}\over{3\pi^2}}
\end{equation}

A necessary condition for bubble nucleation is that $\Delta P>0$. This leads to
an
upper limit on the bag constant, $B_{max}$,
from Eq.~(\ref{deltaP}) as illustrated in Fig.~\ref{neutron}
(the corresponding limit for the Bethe-Johnson equation
of state is shown for comparison; it is seen to be very similar).

Also shown in Fig.\ \ref{neutron} is the limit on the bag constant below which
bubble nucleation takes place at rates exceeding 1 km$^{-3}$Gyr$^{-1}$
and 1 m$^{-3}$s$^{-1}$, respectively, for temperatures of 1, 2, 3 and
10 MeV ($B_{max}$ can be considered as the limit for infinite
temperature). One notes that the possibility of bubble nucleation is
fairly insensitive to the temperature as soon as $T$ exceeds a few MeV,
whereas thermally induced bubble nucleation is impossible for $T <$ 2
MeV (it is known from the stability of ordinary nuclei against decay
into quark matter that $B \geq$ (145 MeV$)^4$). This confirms the
estimate in \cite{deltamu}. The range of bag constants for which
a hot neutron star may transform into quark matter is thus roughly
145 MeV $\leq B^{1/4} \leq$ 152 MeV.

An interesting feature of the solution is the existence of a maximum in
$B$ as a function of $n_B$. This indicates that there is a {\it range in
densities\/} for which boiling can take place for a fixed value of $B$.
If the central density of a neutron star exceeds the upper limit of
$n_B$ permitting boiling, it may therefore happen that boiling is
initiated off-center, with potentially interesting consequences for
supernova energetics, neutrino fluxes, gamma-bursters etc. This may be
explained as follows: With increasing density one has an increase in $\mu$,
and since $P_{ud} \sim \mu^4$ while $P_n \sim \mu^5$, higher densities must
imply still lover $B_{max}$ in order to satisfy the condition, $\Delta P$ =
0. That the maximum does not occur at the same density for all temperatures
is due to the $\mu$-dependence of $\gamma$. This effect,
however, occurs at much higher $n_B$ and $B$ for the more realistic equation
of state discussed below, so we shall not pursue the issue further.

\section{MEAN FIELD APPROXIMATION}
\label{seccompli}

In the following, a mean field model is used to describe the equation
of state in the hadron phase. The model includes the light hadron octet
and is described in detail in Refs.\ \cite{swmodel,omboil,gl91}. For the quark
phase, the only difference from Sec.\ \ref{pure} is that strange quarks are
introduced in accordance with Eq.\ (\ref{matrix}) below. This
leads to additional contributions to Eqs.\ (\ref{deltaP}) and
(\ref{gamma}).

Integrating the Oppenheimer-Volkoff equation gives the structure of the initial
neutron star: Pressure,
chemical potential, and number density of each hadron species as a function
of radius.

For a flavor conserving phase transformation, the relative
number densities, $r_i \equiv n_i/n_B$, are given by

\begin{equation}
\left( \matrix{r_u\cr r_d\cr r_s\cr} \right) = \left(
\matrix{2&1&1&2&1&0&1&0\cr 1&2&1&0&1&2&0&1\cr 0&0&1&1&1&1&2&2\cr} \right)
\left( \matrix{r_p\cr r_n\cr r_{\Lambda}\cr r_{\Sigma^+}\cr r_{\Sigma^o}\cr
r_{\Sigma^-}\cr r_{\Xi^o}\cr r_{\Xi^-}\cr} \right).
\label{matrix}
\end{equation}

The absolute number densities are then given from equality between the
chemical potential per baryon in the two phases.

Doing a calculation similar to the one already
done for a pure neutron star, one obtains a qualitatively similar result
with the limits being independent of the nucleation rate considered
for $T \geq$ 10 MeV but with slightly higher values of $B_{max}$, and with
the maximum in $B$ at higher densities. (Fig.\ \ref{bagcriteos}).

Again, the temperature variation of the $B$-limits show that
there is some critical temperature, $T_{crit} \approx 3 \ {\rm
MeV}$, below which the conversion takes place only for extremely low values
of the bag constant. This corresponds to the limit found by Horvath {\it et
al}.\ \cite{deltamu} although the method for obtaining it is quite different.
As mentioned in the previous section, increasing the
value of $T$ brings the limits closer to the $\Delta P = 0$ - curve.

A typical temperature for a newborn neutron star is about 10 MeV at the
center with off-center temperatures up to 20-30 MeV,
but since we have just seen that the exact value is unimportant in the high
$T$ limit, it is enough
to consider only $T$ = 10 MeV (this agrees well with the choice of $\tau$ =
1 s, since a typical cooling time for newborn neutron stars is of this
order \cite{remark}). The constraints on a
1.4 ${\rm M}_{\odot}$ neutron star is shown on
Fig.\ \ref{bagcritns}. For $B^{1/4} \leq$ 165 MeV the center of the star is
converted, and for lower values of the bag constant still larger fractions
of the star will undergo the phase transition in the initial stage of its
existence. At $B^{1/4}$ = 145 MeV the entire star is transformed but for
this value ordinary nuclei (e.g.\ $^{56}$Fe) would probably be unstable
\cite{farhijaffe}.  For different choices of the hyperon
coupling constants, the limits on $B_{max}$ at the center vary from (162
MeV$)^4$ to (170 MeV$)^4$, with high $B_{max}$ corresponding to low values
of the hyperon-to-hadron coupling constant. Thus a weaker coupling of the
hyperons tends to destabilize hadronic matter.
(The coupling constants mentioned above are chosen to fit the $\Lambda$
binding energy, as described in Ref.\ \cite{gl91}).
At larger radii, the relative density of strange baryons decreases and the
limits converge toward the ones shown in Fig.\ \ref{bagcritns}.

So far, we have only considered $M = 1.4 M_{\odot}$ since observational
data seem to suggest this as the most typical value. The effect of varying the
mass is displayed on Fig.\ \ref{nsmass}. For low masses, the star consists
almost only of neutrons, even at the center, and thus the deviation from
(145 MeV$)^4$ seen here is due only to a nonzero pressure, and it appears
that in this case, a conversion via thermal bubble nucleation is less likely
to take
place. At masses near the maximum mass $B_{max}^{1/4} \rightarrow$ 200 MeV
so here all but unrealistically high $B$ gives a transition into quark
matter.

Turning again to the case $M = 1.4 M_{\odot}$, it could be interesting to
examine the effects of having a non-zero strong coupling constant,
$\alpha_s$. For massless quarks, this gives corrections to the number
density and pressure which to first order is given by $n_q = n_{q,0}
(1-{{2\alpha_s}\over{\pi}})$ and $P_q = P_{q,0} (1-{{2\alpha_s}\over{\pi}})$,
where $0$ denotes the values for $\alpha_s = 0$.

Unfortunately, the corresponding expression for $\gamma (\alpha_s)$ (and in
the case of massive quarks also $\sigma (\alpha_s)$) is presently unknown
and thus we cannot correctly estimate the effect on the nucleation rate.
What we can do, however, is to examine the $\Delta P = 0$-curve, and
thereby obtain limits on $\cal R$, since this is bound to be below the curve
of equal pressure.

It is seen that although both $B_{crit}$ and the limiting bag constant for
$ud$ quark matter stability ($B_{crit}$ taken at $r = R$ corresponding to
zero external pressure \cite{bud}) are decreasing functions of $\alpha_s$, one
has a narrowing of
the relevant interval in $B$ (Fig.\ \ref{alpha}), and above $\alpha_s
\approx 0.6$ conversion of neutron stars seems to be ruled out. (Similar
results were obtained in Ref.\ \cite{kriv}).

The effect on $T_{crit}$ can only be guessed at, but assuming that $W_c
\sim (1-{{2\alpha_s}\over{\pi}})^a$, where $|a| \leq 1-2$, and
$\alpha_s \leq 0.6$, $T_{crit}$ should be correct within a factor of 2, and
thus the temperatures accompanying supernova explosions should still be
enough to ensure conversion into strange stars (or hybrid stars) provided
that the bag constant is below $B_{crit}$.

\section{DISCUSSION AND CONCLUSION}
\label{secclu}

What we have seen is that if the bag constant lies in the interval where
three flavor but not two flavor quark matter is stable at zero
pressure and temperature (145 MeV $\leq B^{1/4} \leq$ 163 MeV, see Ref.\
\cite{farhijaffe}) then all or parts of a neutron star will be converted
into strange matter during the first seconds of its existence (but note the
cautionary remark in \cite{remark}). The rest
will then be transformed either by a slow burning on a time scale of a few
seconds to a few minutes \cite{olinto} or by a detonation \cite{horvath}.
For bag constants above the stability interval, we have seen that a partial
transformation is still possible, but since this seems to depend heavily on
the exact equation of state, one should be careful before drawing any
definite conclusions.

Since a large fraction of the star is converted
on a relatively short time scale, the released energy may well provide a
significant contribution to the total energy of a supernova (cf.\
Ref.\ \cite{horvath}).

Another investigation by Krivoruchenko and Martemyanov \cite{kriv}, taking
$\Delta P = 0$ as a criterion for a possible transformation into strange stars
by a flavor conserving phase transition, have found similar results. This
is an effect not as much of equivalence of the methods used, but rather of
the fact, that the high temperatures of newborn neutron stars together with
the exponential in Eq.\ (\ref{prob}) causes the rate to be insensitive to
$T$.

Another interesting feature is that if a star is born with a
mass that for a given bag constant is too small for the conversion to take
place even in the center, then accretion from a neighboring star, leading
to a higher mass for the neutron star and thus in principle a larger transition
probability, will only lead to a phase transition via thermal nucleation
{\it if} at the same time
the neutron matter is heated to at least 2-5 MeV by the energy released by
the accretion process or by other mechanisms, such as capture of high
energy neutrinos. (It is very unlikely that a sig\-ni\-fi\-cant mass can be
transferred during the first second or so). Thus, one may conclude that if
no mechanism for a significant heating of the star can be found, the
initial mass uniquely determines the future of the star if one has to rely
solely on thermal nucleation. (Other possible mechanisms that may lead to a
transformation were mentioned in Sec.\ \ref{secintro}).

By introducing a non-zero $\alpha_s$, a narrowing in the interesting range
for $B$ was seen; both as an absolute measure and in terms of the fraction
of the interval where $uds$ quark matter is stable at zero pressure.

In this work we have ignored the effect of the mass of the strange quark,
which corresponds to a somewhat
inadequate treatment of the equation of state, surface and curvature effects.
However, even in the center of the star
no more than 2-4 \% of the quarks in the hadrons
are $s$ quarks, and thus the effects during bubble nucleation are very small.
As far as the surface energy is concerned, a more important effect comes
from taking $\sigma_{hadron} \simeq (30 {\rm MeV})^3$ (from typical nuclear
mass formulae), but even here it turns out that
$4\pi r_c^2 \sigma \ll 8\pi r_c \gamma$, so that inclusion of such a term
would not change our conclusions.

\section{ACKNOWLEDGEMENT}

This work was supported by the Danish Research Academy
and the Danish Natural Science Research Council.

\figure{The upper limits on the bag constant allowing boiling of a
nonrelativistic neutron gas into $ud$ quark matter as a function of the
baryon number density in the hadron phase is shown for different nucleation
rates and temperatures. The upper curve for each temperature corresponds to
a rate of one nucleation per ${\rm km}^3$ per Gyr; the lower to one per
${\rm m}^3$ per second. As comparison is shown the $\Delta P = 0$ line for
a Bethe-Johnson equation of state (BJ). This is seen to deviate
only for very large densities.
\label{neutron}}

\figure{Limits on the bag constant in the mean field approximation.
Notation as in Fig.\ \ref{neutron}.
(The model used correspond to x=0.6 in Ref.\ \cite{gl91}).
\label{bagcriteos}}

\figure{Limits on the bag constant for a 1.4 $M_{\odot}$ neutron star.
Notation as in Fig.\ \ref{neutron}.
\label{bagcritns}}

\figure{Critical bag constants in the center of neutron stars as a function
of stellar mass. Notation as in Fig.\ \ref{neutron}.
\label{nsmass}}

\figure{Curves representing pressure equilibrium between the two phases for
various values of the strong coupling constant, $\alpha_s$.
\label{alpha}}


\begin{references}
\bibitem{hoenenogaegget} One should note that the seed-model assumes the
existence of at least one lump of quark matter left over from the Big Bang.
If for some reason quark matter is not formed or is unable to survive the
first second, thermal nucleation or one of the other mechanisms
mentioned in Sec.\ \ref{secintro} is needed to create the first strange
star.
\bibitem{prefactor}
A comment on the $T^4$ prefactor in Eq.\ (\ref{prob}) may be appropriate.
It can be seen as the product of a density and a growth factor, although the
form used here is to a large extent chosen to get the dimension right.
However, as discussed by Olesen and Madsen \cite{omboil}, the dominant factor
is the exponential and substituting e.g. $\mu^4$ for $T^4$ will have only
marginal effects.
\bibitem{remark} However, one should bear in mind that the equation of
state of a supernova core is somewhat different from the model used here
due to the presence of trapped neutrinos.
\bibitem{bud} Strictly speaking, the bag constant where $ud$ quark matter
is marginally stable is only approximately equal to $B_{crit}(r=R)$. The
difference is, however, so small that for this purpose it can be
neglected. The proper limit for $ud$ quark matter stability relative to
iron is $B^{1/4} = 145 {\rm MeV} (1-{{2\alpha_s}\over{\pi}})^{1/4}$. The
corresponding limit for $uds$ stability is
$B^{1/4} = 163 {\rm MeV} (1-{{2\alpha_s}\over{\pi}})^{1/4}$.
\end{references}
\end{document}